\documentclass{vldb}
\usepackage{graphicx}
\usepackage{balance}  
\newcommand{\eat}[1]{}

\usepackage{latexsym}
\usepackage{amsfonts}
\usepackage{amsmath}
\usepackage{amssymb}
\usepackage{color}
\usepackage{colortbl}
\usepackage{epsfig}
\usepackage{xspace}
\usepackage{graphicx}
\usepackage{subfigure}
\usepackage{paralist}
\usepackage{enumerate}
\usepackage[table]{xcolor}
\usepackage[color,matrix,arrow,all]{xy}
\usepackage{cite}
\usepackage{comment}
\usepackage{booktabs}
\usepackage{balance}
\usepackage{stmaryrd}
\usepackage{pifont}
\usepackage{hhline}
\usepackage{listings}
\usepackage{array}
\usepackage{float}
\newcolumntype{L}[1]{>{\raggedright\let\newline\\\arraybackslash\hspace{0pt}}m{#1}}
\newcolumntype{C}[1]{>{\centering\let\newline\\\arraybackslash\hspace{0pt}}m{#1}}
\newcolumntype{R}[1]{>{\raggedleft\let\newline\\\arraybackslash\hspace{0pt}}m{#1}}

\usepackage{multirow}
\usepackage{url}

\usepackage[ruled,vlined,linesnumbered]{algorithm2e}
\usepackage[export]{adjustbox}

\newcommand{\at}[1]{\protect\ensuremath{\mathsf{#1}}\xspace}

\newcommand{\bi}{\begin{itemize}}
\newcommand{\ei}{\end{itemize}}

\newcommand{\be}{\begin{enumerate}}
\newcommand{\ee}{\end{enumerate}}
\newcommand{\beqn}{\begin{eqnarray*}}
\newcommand{\eeqn}{\end{eqnarray*}}

\newcommand{\stitle}[1]{\vspace{1ex}\noindent{\bf #1}}

\newcommand{\etitle}[1]{\vspace{0.8ex}\noindent{\underline{\em #1}}}

\newcommand{\ie}{{\em i.e.,}\xspace}
\newcommand{\eg}{{\em e.g.,}\xspace}

\newcommand{\aka}{\emph{a.k.a.}\xspace}

\newcounter{ccc}

\newcommand{\eop}{\hspace*{\fill}\mbox{$\Box$}\vspace{1ex}}     

\newcommand{\nthesection}{\arabic{section}}

\newcounter{alg}[section]
\renewcommand{\thealg}{\nthesection.\arabic{alg}}

\newcounter{arule}
\renewcommand{\thearule}{\arabic{arule}}

\newcounter{claim}
\renewcommand{\theclaim}{\arabic{claim}}

\newcommand{\sys}{{\sc tac}\xspace}

\makeatletter
    \newcommand\figcaption{\def\@captype{figure}\caption}
    \newcommand\tabcaption{\def\@captype{table}\caption}
\makeatother

\definecolor{shadecolor}{RGB}{200,200,200}

\definecolor{shadecolor1}{RGB}{230,230,230}

\definecolor{shadecolor1}{RGB}{255, 114, 118}

\makeatletter
\def\@copyrightspace{\relax}
\makeatother

\begin{document}

\title{Data Curation with Deep Learning [Vision]}

\numberofauthors{1} 

\author{
	Saravanan Thirumuruganathan~~~~~~Nan Tang~~~~~~Mourad Ouzzani~~~~~~AnHai Doan$^\ddag$\\
	\affaddr{Qatar Computing Research Institute, HBKU~~~~~~~$^\ddag$University of Wisconsin-Madison}
	\\	
	\{sthirumuruganathan, ntang, mouzzani\}@hbku.edu.qa, anhai@cs.wisc.edu
}

\maketitle

\begin{abstract}
Data curation  -- the process of discovering, integrating, and cleaning data -- is one of the oldest, hardest, yet inevitable data management problems.
Despite decades of efforts from both researchers and practitioners, it is still one of the most time consuming and least enjoyable work of data scientists.
In most organizations, data curation plays an important role so as to fully unlock the value of big data. 
Unfortunately, the current solutions are not keeping up with the ever-changing data ecosystem, because they often require substantially high human cost. 
Meanwhile, deep learning is making strides in achieving remarkable successes in multiple areas, such as image recognition, natural language processing, and speech recognition.
%
%
In this vision paper, we explore how some of the fundamental innovations in deep learning could be leveraged to improve existing data curation solutions and to help build new ones.
In particular, we provide a thorough overview of the current deep learning landscape, and identify interesting research opportunities and dispel common myths. 
We hope that the synthesis of these important domains will unleash a series of 
research activities that will  lead to  significantly improved solutions for many data curation tasks.
\end{abstract}

\section{Introduction}
\label{sec:introduction}

\begin{figure*}[!t]
	\centering
	\includegraphics[width=0.94\textwidth]{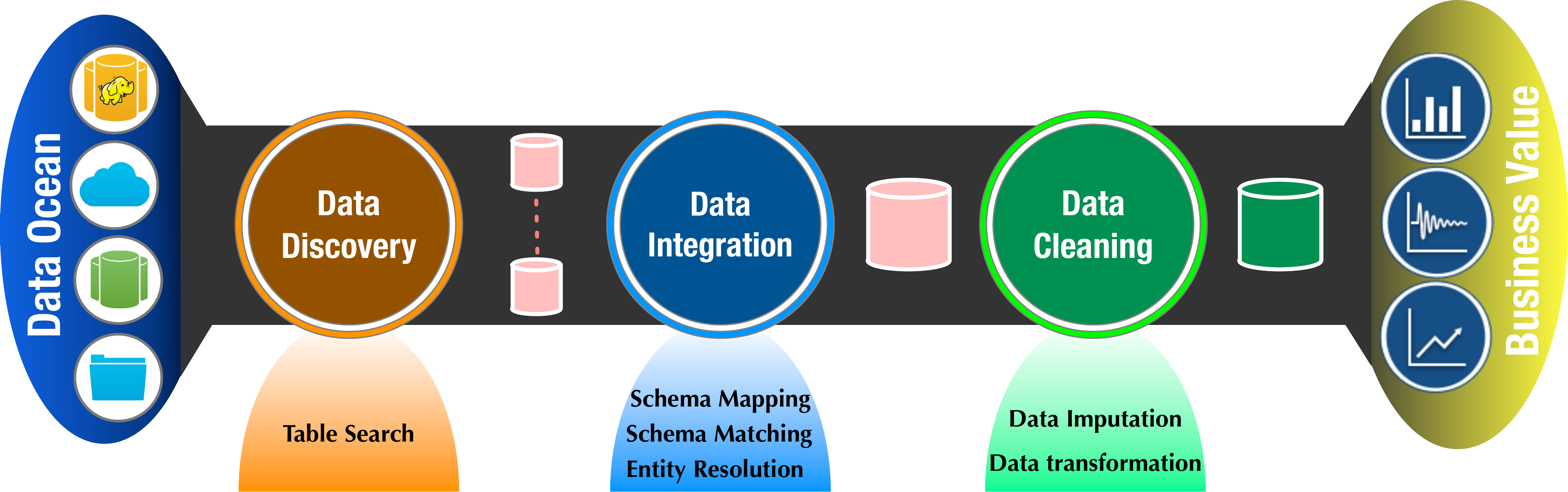}
	\caption{A Data Curation Pipeline}
	\label{fig:dc}
	\vspace{-2ex}
\end{figure*}

\stitle{Data Curation}
-- the process of discovering, integrating and cleaning data for a specific analytics task, as shown in Figure~\ref{fig:dc} -- is critical for 
any organization to extract real business value from their data; feeding flawed, redundant or incomplete data as input will produce nonsense output or ``garbage'' (\aka garbage in, garbage out).
Due to its importance, there has been many commercial solutions (for example, Tamr~\cite{tamr} and Trifacta~\cite{DBLP:conf/cidr/HeerHK15}) and academic  efforts for all aspects of DC, including data discovery~\cite{civilizer,sspaper,DBLP:journals/debu/MillerNZCPA18,DBLP:journals/pvldb/NargesianZPM18}, data integration~\cite{DBLP:journals/pvldb/KondaDCDABLPZNP16,dataintegration,swoosh},
and data cleaning~\cite{dqm,nadeef,bigdansing,cfd,fixingrule,holoclean,DBLP:conf/vldb/RamanH01}.

Unfortunately, DC remains the most expensive task for humans -- an oft-cited statistic is that data scientists spend 80\% of their time curating their data~\cite{civilizer}.
The reason for such a high human cost is that most DC solutions cannot be fully automated, as they are often {\em ad-hoc} and 
require substantial effort to generate things, such as features and labeled data, 
which are used to synthesize rules and train machine learning models.
Existing solutions are not keeping up with the ever-changing data ecosystem in terms of volume, velocity, variety and veracity (\aka the four V's of big data) -- practitioners are desperate for 
practical and usable solutions that can significantly reduce the human cost.

\stitle{Deep Learning}
is an emerging paradigm within the area of machine learning (ML) that has achieved incredible successes in diverse areas, such as computer vision, natural language processing, speech recognition, genomics, and many more.
The trifecta of big data, better algorithms, and faster processing power has resulted in DL achieving super human performance in many areas.
Due to these successes, there has been extensive new research seeking to apply DL to other areas both inside and outside of computer science.

\stitle{Data Curation Meets Deep Learning.} 
This article investigates intriguing research opportunities for answering the following questions:

\bi
	\item What does it take to significantly advance a challenging area such as DC?
	\item How can we leverage techniques from DL for DC?
	\item Given the many DL research efforts, how can we identify the most promising leads that are most relevant to DC?
\ei

We believe that DL brings unique opportunities for DC, a critical task in data management, which our community must seize upon.
In this article, we lay down a roadmap on how to make this a reality.
Based on a thorough analysis, we identified three fundamental ideas that have a great potential to solve challenging DC problems.


\etitle{Feature (or Representation) Learning.}
For either ML- or non-ML based DC solutions, domain experts are heavily involved in 
feature engineering and data understanding so as to define data quality rules or make sophisticated deductive inference.
One of the fundamental ideas in DL is \emph{representation learning} 
where appropriate features for a given task are automatically learned instead of being manually crafted. 
Developing DC-specific representation learning algorithms could dramatically alleviate many of the frustrations domain experts face when solving DC problems.
Ideally, such representation must be learned in an unsupervised manner to minimize the human labor.
We should be able to learn an appropriate representation from a wide variety of data sources 
that are often scattered across various modalities including structured data (e.g., relations), 
unstructured data (e.g., documents), graphical data (e.g., enterprise networks), and even videos, audios, and images.
The learned representation must be generic such that it could be used for multiple DC tasks. 

\etitle{DL Architectures for DC.}
So far, no DL architecture exists that is cognizant of the characteristics of DC tasks, 
such as representations for tuples or columns, integrity constraints, and so on.
Naively applying existing DL architectures may work well for some, but definitely not all, DC tasks.
In retrospect, computer vision blossomed mainly because of convolutional neural networks (CNNs), and 
natural language processing (NLP) and speech recognition flourished because of sequential DL architectures such as recurrent neural networks (RNNs).
Analogously, we need {\em an equivalent of such a DL architecture customized for DC task(s)}.

\etitle{Taming ML's Hunger for Data.}
While ML methods give state-of-the-art performance for many DC tasks, they also require large amounts of labeled data.
Similarly, the super human performance of DL in various domains are perceived to be on top of a very large amount of training data. 
Hence, naively using DL for DC is doomed to fail as generating high quality training data is expensive and demanding for the domain experts.
However, there has been extensive research in the DL community about how to learn from limited data.
There are a number of promising techniques for a data scarce domain such as DC that we could adapt.

\begin{figure*}[t!]
	\centering
	\includegraphics[width=\textwidth]{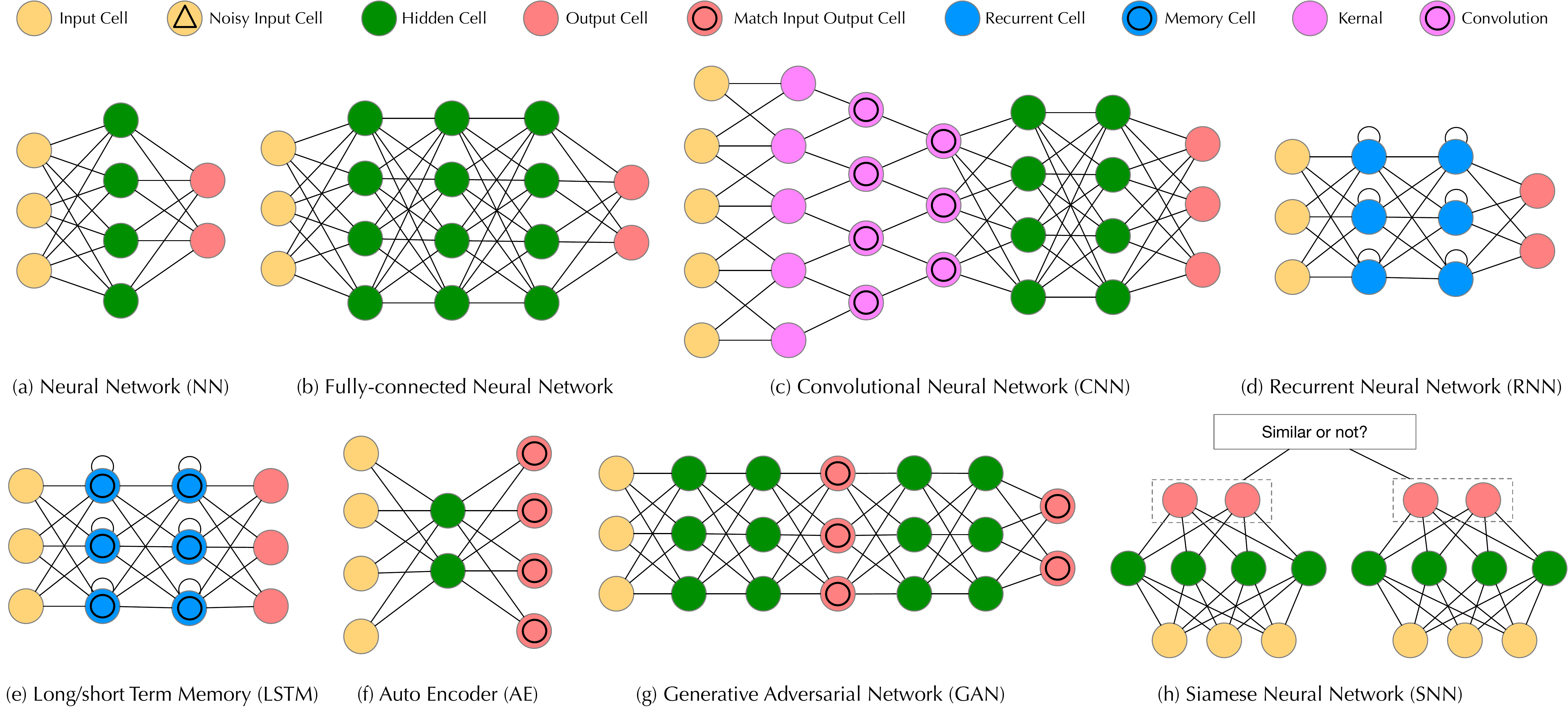}
	\caption{Deep Learning Architectures}
	\label{fig:dla}
\end{figure*}

\stitle{Contributions and Roadmap.}
%
In this paper, we describe the main elements of \sys ({\em Towards Automatic Curation}), an ecosystem 
for our vision to unleash a DL-driven revolution in DC by exploiting the huge potential offered by DL to tackle key DC challenges.
The elements of our vision are listed below:

\begin{itemize}
     \item {\bf Representation Learning for Data Curation.} 
	 	We describe several  research directions for building representations that are explicitly designed for DC. 
		Developing algorithms for representation learning that 
		can work on multiple modalities of data (structured, unstructured, graphical), and
		can work on diverse DC tasks (such as deduplication, error detection, data repair) is challenging (Section~\ref{sec:representationLearning}). 

	\item {\bf Deep Learning for Data Curation.}
		We identify some of the most common tasks in DC and propose preliminary solutions using DL. 
		We also highlight the importance of designing DC specific DL architectures.
		Finally, we briefly describe our early successes in applying DL to some DC problems, 
		such as data discovery~\cite{sspaper}  and entity resolution~\cite{deeper, Mudgal:2018:DLE:3183713.3196926}
		(Section~\ref{sec:DCforDL}).

    \item {\bf Taming \sys's Hunger for Data.}
		DL rightly has a reputation for requiring a large amount of training data.
		We describe a series of techniques (\eg unsupervised representation learning, data augmentation, synthetic data generation, weak supervision, domain adaptation, and crowdsourcing), which when adapted for DC, 
		could dramatically reduce the required amount of labeled data (Section~\ref{sec:dldcHunger}).

	\item {\bf Myths and Concerns about DL.}
    	DL is still a relatively new concept, especially for database applications and is not a silver bullet. 
		We thus identify and address a number of myths and concerns about DL (Section~\ref{sec:discussion}).
    
\end{itemize}
We close this paper by a call to arms (Section~\ref{sec:conclusion}). 
~\\

Since there is a Cambrian explosion of DL research with thousands of papers being published every year that can drown an unguided researcher, we also provide a crash course on DL through a series of carefully selected of concepts that are relevant to DC (Section~\ref{sec:preliminaries}).
%
%
%


\section{Deep Learning Fundamentals}
\label{sec:preliminaries}

This section is a crash course of fundamental DL concepts (see~\cite{Goodfellow-et-al-2016, chollet2018deep, lecun2015deep} for more details) needed in the latter sections. While these concepts are widely discussed in the DL literature, we include them here for completeness -- even a general audience can understand our work.
Of course, readers with DL background can safely skip this section.


{\em Deep learning} is a subfield of ML that seeks to learn meaningful representations from the data that could be used to solve the task at hand effectively.
The most commonly used DL models are neural networks with many hidden layers.
The key insight is that successive layers in this ``deep'' neural network can be used to learn increasingly useful representations of the data.
Intuitively, the input layer often takes in the raw features.
As the data is forwarded and transformed by each layer, more and more meaningful information (representation) is extracted.
This incremental manner in which increasingly sophisticated representations are learned layer-by-layer is one of the defining characteristic of DL.
The number of layers in a model is referred to as its \emph{depth} 
while the number of parameters to be learned is the model's \emph{capacity}.

In the following, we first introduce some DL architectures (Section~\ref{subsec:dla}), which are depicted in Figure~\ref{fig:dla}\footnote{The figure's style is based on the Neural Network Zoo by Fjodor van Veen (\url{http://www.asimovinstitute.org/neural-network-zoo/}).}. 
We then discuss distributed representations (Section~\ref{subsec:dldr}).

\subsection{Deep Learning Architectures}
\label{subsec:dla}

\subsubsection*{(a) Neural Networks}

Neural networks are a biologically inspired ML model that mimic human brains.
Figure~\ref{fig:dla}(a) shows an example of a three layer neural network,
which contains an {\em input} layer (the leftmost layer),
a {\em hidden} layer (the intermediate layer), and
an {\em output} layer (the rightmost layer).
In a nutshell, neural networks are simply layers stacked on top of each other, 
with the neurons in each layer not connected among themselves.
Each layer can be thought of performing a geometric transformation of its input such that the resulting output is more meaningful for the ML task.
That is, the neural network performs a complex geometric transformation (where each layer performs over the output of the other)
of the input data space to the output data space.
The transformation is parameterized by the weights of each layer that are learned by an algorithm based on the training data.
Initially, they are often set to random values corresponding to a random -- and probably useless -- transformation.
The learning algorithm evaluates how far the output of the network is from the training data through \emph{loss scores}.
It then systematically updates the parameters such that the loss score decreases.
This process is often repeated many times to obtain a neural network whose output is as close as possible 
to the output from the training data.

\subsubsection*{(b) Fully-connected Neural Networks}

The most common models used in DL are neural networks with many hidden layers (see Figure~\ref{fig:dla}(b)) --
a series of layers where each node in a given layer is connected to every other node in the next layer.
Fully-connected neural networks are also called feed-forward neural network, 
since they feed information from the input to the output.
It can learn relationships between any two input features or intermediate representations.
Its generality however comes with a cost; one has to learn weights and parameters which requires a lot of training data.
For example, two fully connected hidden layers with 100 nodes in each layer will require learning 10,000 parameters.

\subsubsection*{(c) Convolutional Neural Networks (CNNs)}

CNNs are feed-forward neural networks (see Figure~\ref{fig:dla}(c)).
They differ from other neural networks in that the input is fed through convolutions layers, where neurons in convolutional layers only connect to close neighbors (instead of all neurons connecting to all neurons). Also, convolutional layers tend to shrink as they become deeper (\eg by easily divisible factors of the input).
CNNs are widely used by the computer vision community.
Because every image can essentially be represented as a matrix of pixel values, 
instead of learning global patterns between arbitrary features, this method focuses on 
spatially local patterns.
If a layer can recognize a certain pattern, it can recognize it irrespective of its location in the image.
Furthermore, CNNs also have the ability to learn spatial hierarchies such as {\sl nose} $\rightarrow$ {\sl face} $\rightarrow$ {\sl human}. 
In other words, CNNs can learn increasingly complex (resp. abstract) representations from simpler (resp. granular) representations.

\subsubsection*{(d) Recurrent Neural Networks (RNNs)}

RNNs are also feed-forward neural networks (see Figure~\ref{fig:dla}(d)).
In contrast to the previous DL architectures that process the input on its entirety, RNN processes them one step at a time (for example, given two words ``data curation'', it first handles ``data'' and then ``curation'').
Hence, neurons in an RNN are designed by being fed with information not just from the previous layer but also from themselves from the previous pass. Consequently, the order of feeding an input to RNN matters.
RNNs are widely used in NLP and speech recognition. 

\subsubsection*{(e) Long Short Term Memory (LSTM)}

In practice, a simple RNN often has troubles with learning long range dependencies.
Long Short Term Memory (LSTM) (see Figure~\ref{fig:dla}(e)) is a popular variant of RNN that can be explicitly used when the input sequence exhibits long range dependencies.
LSTM provides a mechanism to ``remember'' past information across multiple time steps through its memory cells.


\subsubsection*{(f) Auto Encoders (AE)}

Autoencoders are a popular DL model for unsupervised learning of efficient representations and used commonly in feature extraction (see Figure~\ref{fig:dla}(f)).
An autoencoder takes a $d$-dimensional input $x$, and maps it to a hidden representation $y$ in a $d^\prime$-dimensional space using \emph{encoders}.
This hidden representation can then be \emph{reconstructed} to a $d$-dimensional output $z$ using a \emph{decoder}.
Of course, for the autoencoder to learn useful representations, 
we require that $z$  be as close to $x$ as possible and that $d^\prime < d$ (otherwise, the network can simply ``memorize'' the input).
This forces the autoencoder to learn a compressed representation of the input.
Intuitively, an autoencoder seeks to reconstruct its own inputs -- given an input $x$, and reconstructs it as accurately as possible 
by encoding the input to a low dimensional latent space 
(that can capture the major structures/patterns implicit in the data) 
and then decoding it back to the original space.

\subsubsection*{(g) Generative Adversarial Networks (GANs)}

GANs~\cite{goodfellow2014generative} are a class of DL models used for unsupervised representation learning.
They work as twins: two neural networks working together -- a \emph{generator} and a \emph{discriminator} (Figure~\ref{fig:dla}(g)), 
where the former generates content that will be then judged by the latter.
Similar to autoencoders, the generator seeks to map an input vector from the latent space to a vector in the data space.
For example, let us consider the domain of (synthetic) generation of photo-realistic images.
The generator tries to map a vector in the latent space to a (synthetic) image.
The discriminator takes an input (either real or synthetic) and predicts whether the image is original or produced by the generator.
The goal of the system is to increase the number of mistakes made by the discriminator -- 
which implies that the data generated are so realistic that they can fool the discriminator.
\cite{chollet2018deep} gives a forger-dealer analogy where a forger seeks to create fake (say) Picasso paintings
while the dealer tries to distinguish between the original paintings and the fakes.
While the forger is initially bad, eventually she becomes increasingly competent at imitating Picasso's style.
Of course, the dealer also becomes increasingly competent at recognizing fakes.
When these competing systems converge, the results are very realistic Picasso paintings that can fool all but very best Picasso experts.

\subsubsection*{(h) Siamese Neural Network (SNN)}

SNN is a class of neural network architectures that contain two (or more) identical subnetworks (see Figure~\ref{fig:dla}(h)). These two subnetworks have the same configuration with the same parameters and weights. Parameter updating is mirrored across both subnetworks.
SNNs are popular among tasks that involve finding similarity or a relationship between two comparable things.

\subsection{Distributed Representations}
\label{subsec:dldr}

\begin{figure}[t!]
	\centering
	\includegraphics[width=\columnwidth]{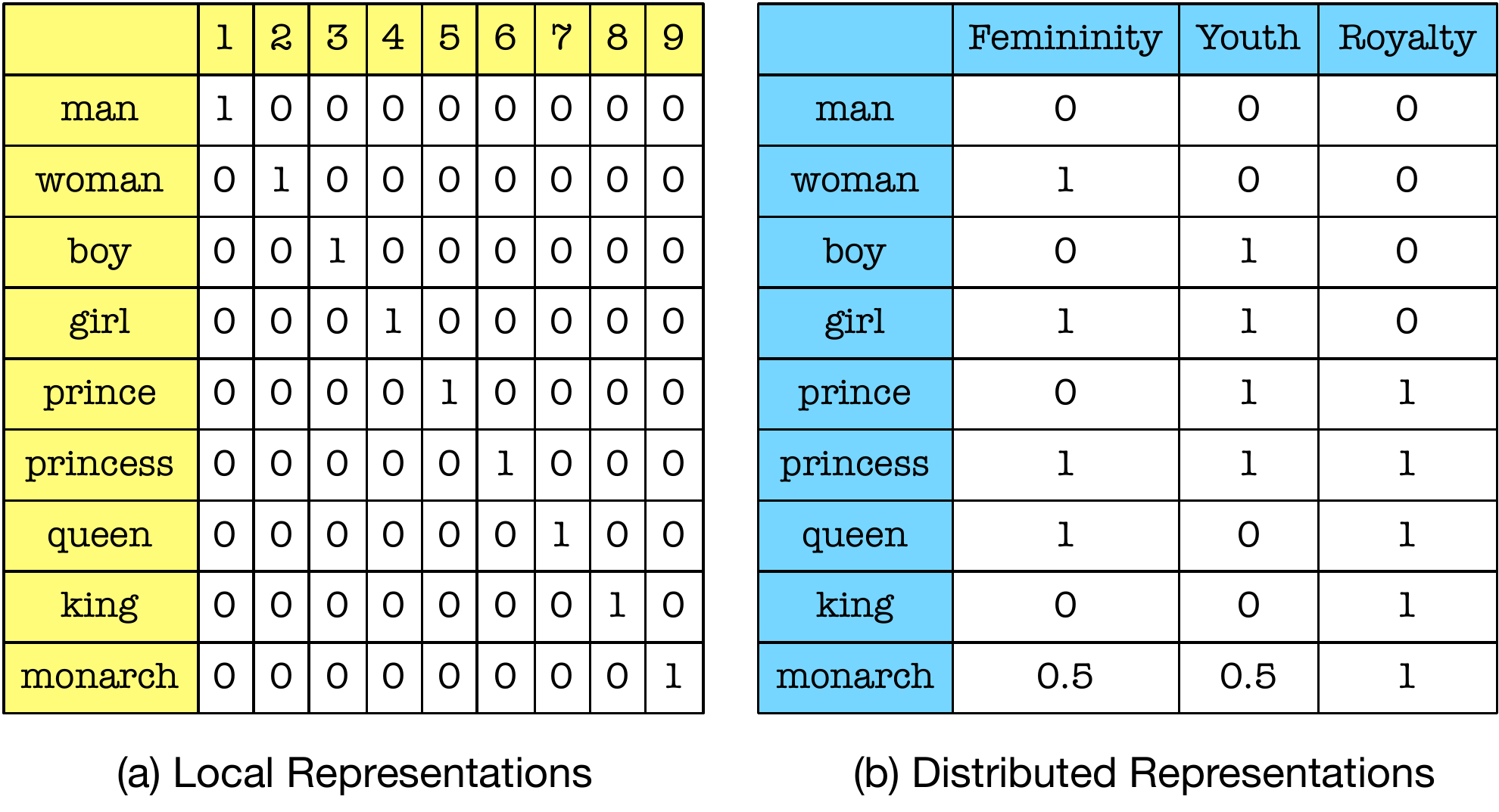}
	\caption{Local vs. Distributed Representations}
	\label{fig:dr}
\end{figure}

Deep learning is based on learning data representations, and the concept of distributed 
representations (\aka embeddings) is often central to DL.
In contrast to {\em local representations} where each object is represented by a single representational element (or one neuron in neural networks), {\em distributed representations}~\cite{hinton1986learning} represent one object by many representational elements (or many neurons).
Conversely, each neuron is associated with more than one represented object.
That is, the neurons represent features of objects.

Let us better illustrate this concept through an example.
Consider Figure~\ref{fig:dr} with different distributed representations for words such as {\em man}, {\em woman}, {\em boy}, and so on\footnote{The example is from \url{https://www.shanelynn.ie/get-busy-with-word-embeddings-introduction/}}.
Local representations are one-hot (or ``$1$-of-$N$'') encodings, where all except one of the values of the vectors are zeros (see Figure~\ref{fig:dr}(a)).
If we try to reduce the dimensionality of the encoding, we can use distributed representations (see Figure~\ref{fig:dr}(b)) where each word is represented by several dimensions with decimal values between 0 and 1.

There are certain advantages of using distributed representations.
(1)~The representation power is exponential in the total dimensions available, while the representation power of local representations is only linear to the total dimensions.
(2)~It is easier to capture semantic similarity between two different objects, \eg~{\sl girl} is closer to {\sl princess} than {\sl man} in two dimensions (\at{Femininity} and \at{Youth}).
These advantages become especially relevant in domains such as DC.

\subsubsection{Distributed Representations of Words}

Distributed representations of words (\aka word embeddings) seek to map individual words to a vector space, which helps learning algorithms to achieve better performance in NLP tasks by grouping similar words. 
For word embeddings, the dimensionality is often fixed (such as $300$) and the representation is often \emph{dense}.
Moreover, the representation is \emph{distributed} where each word is represented by multiple components of the representations (all the non-zero values of the vector)
and each component can potentially participate in representing multiple words 
(\ie the $i$-th component can take a non-zero value and can be used in the representation of many words).

Word embeddings are often learned from the data in such a way that semantically related words are often close to each other.
In other words, the geometric relationship between words often also encodes a semantic relationship between them.
Furthermore, it is also possible that certain vector operations have a semantic interpretation.
A oft-quoted example shows that by adding the vector corresponding to the concept of {\sl female} to the distributed representation of {\sl king},
we (approximately) obtain the distributed representation of {\sl queen}.
Popular word embeddings such as word2vec~\cite{word2vec} often encode hundreds if not thousands of such meaningful transformation vectors, 
such as singular-to-plural, symptom-disease, country-capital, and so on.

\subsubsection{Distributed Representations of Graphs}

Distributed representations of \emph{graphs} (\aka network embeddings) are a popular mechanism to effectively learn appropriate features for graphs (see~\cite{networkembedding} for a survey).
Typically, one seeks to represent each node in the graph as a dense fixed length high dimensional vector.
The optimization objective is often neighborhood-preserving whereby two nodes that are in the same neighborhood will have similar representations.
Depending on the definition of neighborhood, one can design different representations for nodes in a graph.

\subsubsection{Compositions of Distributed Representations}

Most of the distributed representation learning algorithms are designed for atomic units -- words in NLP, or nodes in a graph.
One can then use these to design distributional representations for more complex units.
In the case of NLP, these could be sentences (\ie sentence2vec), paragraphs or even documents (\ie doc2vec)~\cite{doc2vec}.
In the case of graphs, it can be subgraphs or entire graph.
Analogously, for databases, the atomic unit is a cell (an attribute value of a tuple).
Assuming that we can learn the distributed representations of cells,
by composition, we can design representations for tuples, columns, tables, or even an entire database.

Recently, there has been an increasing interest in \emph{directly} learning the distributed representations based on the task at hand (\ie bypassing the composition phase).
For example, one can directly learn distributed representations of subgraphs~\cite{narayanan2016subgraph2vec} and use it for community detection.

\section{Representation Learning for  Data Curation}
\label{sec:representationLearning}

On the one hand, one of the major pain points in DC is feature engineering (as discussed earlier in Section~\ref{sec:introduction}), because this process is manual, time consuming and often requires substantial domain knowledge.
On the other hand, the success of DL, such as in the fields of computer vision and NLP, is largely due to the ability to automatically learn appropriate features/representations.


Hence, a major research focus is {\em how to (automatically) learn representations} that can serve multiple DC tasks. 
Unfortunately, directly applying representation learning algorithms designed for other domains may be inadequate or infeasible.
However, the bright side is that it provides a good opportunity to investigate DC specific representation learning, which will be discussed in this section.


The rest of this section is organized as follows.
Section~\ref{subsec:drdc} discusses how various seemingly disparate DC tasks are connected to distributed representations through the notion of matching.
Next, Section~\ref{subsec:dlCell} proposes an approach to learn distributed representation for cells 
that synthesizes ideas from NLP (word embeddings) and graphs (node embeddings).
Finally, Section~\ref{subsec:dlCompositional} discusses how to leverage 
two emerging ideas from NLP -- hierarchical and contextual representations -- for DC. 

\subsection{Distributed Representations} 
\label{subsec:drdc}

The ``matching'' process is a central concept in most, if not all, DC problems, more specifically,
\bi
	\itemsep0em
	\item {\em data discovery:} find datasets that {\em match} a user specification (\eg a keyword or an SQL query);
	\item {\em schema mapping:} capture {\em matched} columns, \eg those from the same domain;
	\item {\em entity resolution:} discover {\em matching} entities (or tuples), \ie those refer to the same real-world objects;
	\item {\em outlier detection:} detect anomalous data that {\em does not match} a group of values, either syntactically, semantically, or statistically;
	\item {\em data imputation:} guess the missing value that should {\em match} its surrounding evidence; and
	\item {\em data dependencies:} know whether one attribute value depends on other attribute values, which is widely used as integrity constraints to capture data violation -- this could be treated as {\em relevance matching}.
\ei

The main drawback of traditional DC solutions is that they typically use ``local'' interpretation for each DC task as mentioned above. Consequently, these solutions are often ad-hoc and thus not extensible.
Intuitively, distributed representations, which can interpret one object in many  different ways, may provide great help for various DC problems. 
However, there are a number of challenges before they could be used in DC.
In domains such as NLP, simple co-occurrence is an sufficient approximation for distributional similarity.
In contrast, domains such as DC require much more complex syntactic and semantic relationships between (multiple) attribute values, tuples, columns, and tables.
Furthermore, the data itself could be noisy and the learned distributions should be resilient to errors.

\subsection{Distributed Representation of Cells}
\label{subsec:dlCell}

A {\em cell}, which is an attribute value of a tuple, is the smallest data element in a relational database.
As we shall show now, learning distributed representation for cells is already quite challenging and 
requires synthesis of representation learning of words and graphs. 

\subsubsection{An Approach Adapted from Word Embeddings}

A plausible approach inspired by word2vec~\cite{mikolov2013distributed} treats this as equivalent to the problem of learning word embeddings, where we map words to a dense high dimensional vector such that semantically related words are close to each other (see more details in Section~\ref{subsec:dldr}).
A naive adaptation treats each tuple as a document where the value of each attribute corresponds to words. 
This setting will ensure that if two attribute values occur together often in a similar context, then their distributed representation will be similar.
For example, if a relation has attributes \at{Country} and \at{Capital} with many tuples containing {\sl (USA, Washington DC)}, then their distributed representations would be similar.

\stitle{Limitations of this approach.}
There are several obvious limitations when simply applying the method of learning word embeddings for learning cell embeddings. These include:

\begin{itemize}
	\itemsep0em
	\item Databases are typically well normalized to reduce  redundancy, \eg a databases in third normal form (3NF) or Boyce-Codd normal form (BCNF)~\cite{AbiteboulHV95}, which also minimizes the frequency that two semantically related attribute values co-occur in the same tuples.

	\item The window size $W$, which is used  to scan the document by considering a  word and a window of words around it, may have a dramatic impact on learning cell embeddings. 
	Consider a relation $R(A_1, \ldots, A_m)$, where $A_i$ is \at{Country} and $A_j$ is \at{Captial} that are clearly relevant to each other. 
Since tuples are treated as a document and thus some order is assumed, if $|i - j| = 1$, a window size of $W = 2$ can capture their co-occurrences.
However, if $|i - j| > k$ ($k$ is a relatively large number, say 10), then even a window size $W = 10$ will miss them.

	\item A big difference between databases and documents is that databases have many data dependencies (or integrity constraints), within tables (\eg functional dependencies, and conditional functional dependencies~\cite{cfd}) or across tables (\eg foreign keys, and matching dependencies~\cite{DBLP:journals/pvldb/FanJLM09}). These data dependencies are important hints about  semantically related cells that should be captured by learning distributed representations of cells.
\end{itemize}

\subsubsection{Combining Word and Graph Embeddings}

In order to also capture the relationships (\eg integrity constraints) between cells, a more natural way is to treat each relation as a {\em heterogeneous network}.
More specifically, each relation $D$ is modeled as a graph $G(V, E)$,
where each node $u\in V$ is a unique attribute value, and each edge $(u, v) \in E$ represents multiple relationships, such as $(u, v)$ co-occur in one tuple, there is functional dependency from the attribute of $u$ to the attribute of $v$, and so on. The edges could be either directed or undirected, and may carry different labels and weights.
Intuitively, using this enriched model might provide a more meaningful distributed representation that is cognizant of both content and constraints.

\begin{figure}[!t]
	\centering
	\includegraphics[width=.839\columnwidth]{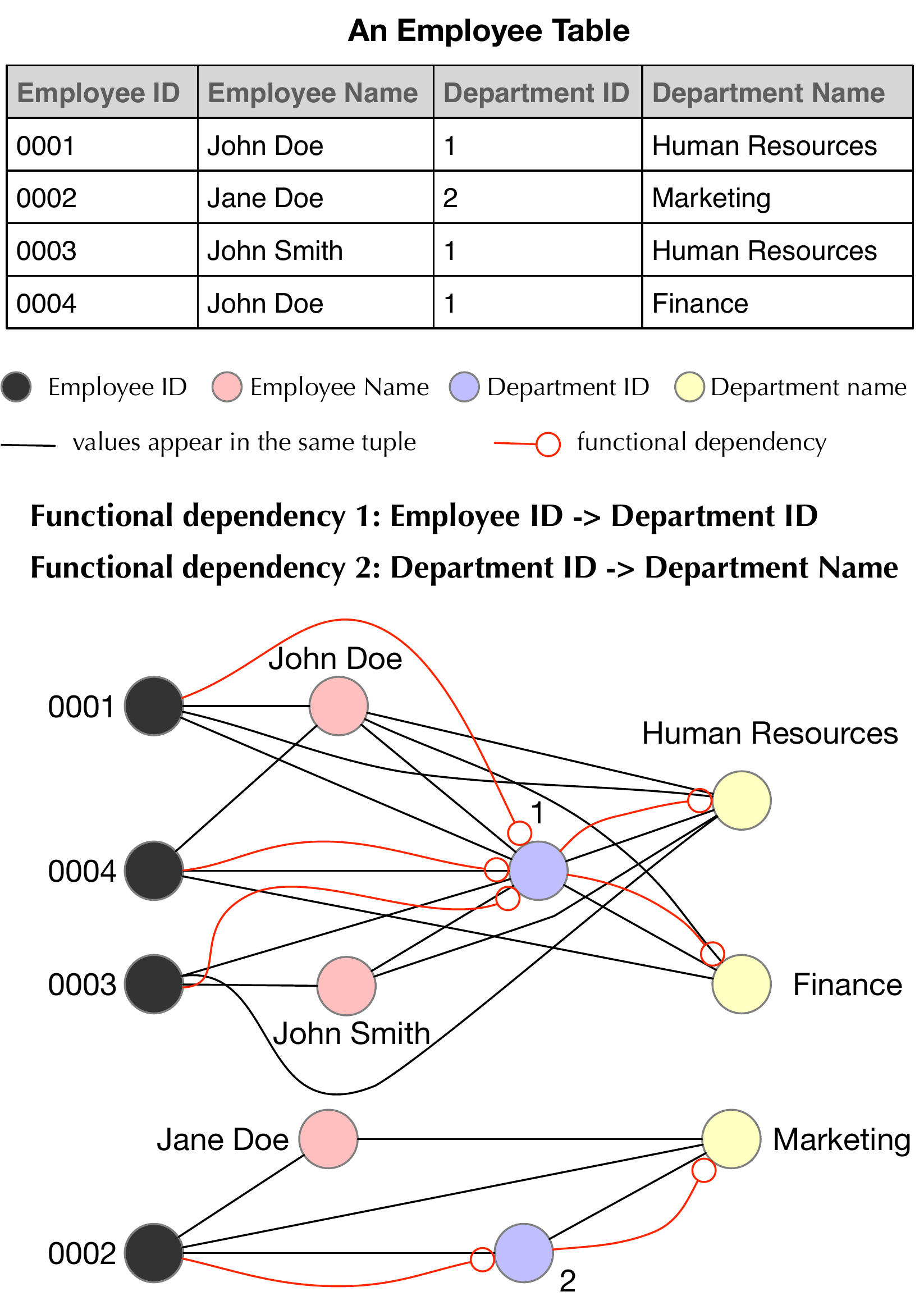}
	\caption{A Heterogeneous Graph of A Table}
	\label{fig:drcell}
\end{figure}

A sample table and our proposed graph representation of the table is shown in Figure~\ref{fig:drcell}. There are four distinct \at{Employee~ID} values (nodes), three distinct \at{Employee~Name} values, two distinct \at{Department~ID} values, and three \at{Department~Name} values. There are two types of edges: undirected edges indicating values appearing in the same tuple, \eg 0001 and John Doe, 
and directed edges for functional dependencies, \eg \at{Employee~ID} 0001 implies \at{Department ID} 1.

\subsubsection*{Research Opportunities}

\begin{itemize}
	\itemsep-1ex
    \item {\em Algorithms.} 
		It is desirable to learn a representation that takes both content and constraints into account. 
		In general, how can we design an algorithm for learning cell embeddings that take 
		values, integrity constraints, and other metadata (\eg a query workload) into consideration?
    
    \item {\em Global Distributed Representations.} 
    	We need to learn distributed representations for the cells over the entire data ocean, not only on one relation. 
		How can we ``transfer'' knowledge gained from one relation to another one such that the learned representations are more meaningful and powerful?
    
    \item {\em Handling Rare Values.} 
    	Rare values, such as primary keys, should be treated fairly, so as to have meaningful representations.
\end{itemize}

\subsection{Hierarchical and Contextual Distributed Representations}
\label{subsec:dlCompositional}

The idea proposed in the previous section seeks to learn the distributed representations for cells.
However, many DC tasks are often performed at a high level of granularity.
The next fundamental question to solve is to design an algorithm to \emph{compose} the distributed representations of more abstract units from these atomic units.
As an example, how can one design an algorithm for tuple embeddings assuming one already has a distributed representation for each of its attribute values?
A common approach is to simply \emph{average} the distributed representation of all its component values.
Alternatively, one can use a more sophisticated approach such as LSTM that follows a data driven approach to compose the tuple's distribution while taking into account 
long range dependencies inherent to a relation.

\subsubsection*{Research Opportunities}
\begin{itemize}
	\itemsep-1mm	
    \item {\em Tuple Embeddings (Tuple2Vec)}: 
		Is there any other elegant approach to compose representations for tuples? 
		Can it be composed from representations for cells? Or should it be learned directly?

    \item {\em Column Embeddings (Column2Vec)}: 
		Many tasks such as schema  matching require the ability to represent an entire column (\ie  attribute) as a distributed representation. 
        How can one adapt the ideas described above for columns?

    \item {\em Table Embeddings (Table2Vec) or Database Embeddings (Database2Vec)}:
     	Tasks such as copy detection or data discovery (finding similar relations) 
        may require to represent an entire relation or even an entire database as a single vector. 
		How can one design embeddings for tables or databases? 
        What are the research challenges in designing such a representation?

    \item  {\em Direct Learning of Hierarchical Representations}: 
		There has been a series of proposals that seek to directly learn representations for composite objects, such as 
        sentences and paragraphs~\cite{le2014distributed}, graphs~\cite{narayanan2017graph2vec}, and subgraphs~\cite{narayanan2016subgraph2vec}. 
        An interesting direction is to adapt these ideas to directly learn the representations for tuples, columns, tables and databases.

	\item {\em Compositional or Direct Learning}:
		Different domains require different ways to create hierarchical representations.
		For the computer vision domain, hierarchical representations such as shapes are learned by composing simpler individual representations such as edges.
		On the other hand, for NLP, hierarchical representations are often learned directly~\cite{le2014distributed}.
		What is an appropriate mechanism for DC?

	\item {\em Contextual Embeddings for DC.}
		Recently, there has been increasing interest in the NLP community in contextual embeddings~\cite{peters2018deep,devlin2018bert}.
		The key insight is that naive word embeddings assign the same vector representation for apple (the fruit) and Apple (the company).
		By assigning the embedding based on the context, it is possible to improve the performance on various tasks that require word disambiguation.
		Often, DC tasks  require a number of contextual information and hence any learned representation must take that into account.
		It is an important challenge to come up with appropriate formalization of context and algorithms for contextual embeddings in DC. 

	\item {\em Multi Modal Embeddings for DC.}
		Enterprises often possess data across various modalities, such as structured data (\eg relations), unstructured data (\eg documents), 
		graphical data (\eg enterprise networks), and even videos, audios, and images.
		Moreover, most of the prior work in DC treats various modalities of data separately, 
		\eg keys for graphs~\cite{DBLP:journals/pvldb/FanFTD15} have nothing to do with keys for tables, even if the information stored in graphs and tables are relevant.
		An intriguing line of research is \emph{cross modal representation learning}~\cite{DBLP:conf/ijcai/HuangPY17}, 
		wherein two entities that are similar in one or more modalities, 
		\eg occurring in the same relation, document, image, and so on, will have similar distributed representations.
\end{itemize}

\section{Deep Learning Architectures for Data Curation Tasks}
\label{sec:DCforDL}

Representation learning and domain specific architectures are the two pillars that led to the major successes achieved by DL in various domains. 
While representation learning ensures that the input to the DL model contains all relevant information,
the domain specific architecture often processes the input in a meaningful way requiring less training data than generic architectures.
As we shall discuss later, the popular DL architectures for computer vision (such as CNN) are very different from those for NLP (such as RNN). 

We organize the section as follows.
We begin by motivating the need for  DL architectures specifically designed for DC in Section~\ref{subsec:dlArchForDC}. 
Next, we discuss how some of the recent innovations in language modeling could guide us in designing such architectures in Section~\ref{subsec:dlDCArchLangModeling}.
We switch gears in Section~\ref{subsec:preTrainedModels} to show these specialized models 
could be trained on large corpus of data such that they can be used by other organizations.
For example, models such as Word2vec from Google has been widely used by many organizations for solving their NLP tasks.
Finally, we briefly discuss some of our early successes from QCRI and UW-Madison on 
designing DL architectures for specific DC tasks such as data discovery and entity matching in Section~\ref{subsec:lowHanging}.

\subsection{Need for DC Specific DL Architectures}
\label{subsec:dlArchForDC}

Recall from Section~\ref{sec:preliminaries} that a fully connected architecture (Figure~\ref{fig:dla}(b)) is the most generic one;
It does not make any domain specific assumptions and hence can be used for arbitrary domains.
However, this generality comes at a cost: a lot of training data.
One can argue that a major part of DL's success in computer vision and NLP is due to the design of specialized DL architectures -- CNN and RNN respectively.
For example, CNN leverages the fact that image recognition tasks often exhibit 
spatial hierarchies of patterns where complex/global patterns are often composed of simpler/local patterns (\eg~{\sl curves $\rightarrow$ mouth $\rightarrow$ face}).
Similarly, RNN assumes that language often requires sequential processing.
In other words, it processes an input sequence one step at a time while maintaining an internal state.
This is analogous to how humans process a sentence -- one word at a time while maintaining an internal state of the sentence's meaning based on what we have read so far.
When we complete the entire sentence, we use this internal state to comprehend its meaning.
These assumptions allows one to design effective neural architectures for processing images and text that also require less training data.

There is a pressing requirement for new DL architectures tailored for DC that are cognizant of the characteristics of DC tasks.
They can then be leveraged for efficient learning in terms of training time and the amount of required training data.

\subsubsection*{Desiderata for DC-DL Architectures}

\begin{itemize}
    \item {\em Handling Heterogeneity.}
		In both CNN and RNN, the input is homogeneous -- images and text. 
		However, a relation can contain a wide variety of data, such as categorical, ordinal, numerical, textual, and image. 

	\item {\em Set/Bag Semantics.}
		While an image can be considered as a sequence of pixels and a document as a sequence of words,
		such an abstraction does not work for relational data.
		Furthermore, the relational algebra is based on set and bag semantics with the major query languages specified as set operators. 
		DL architectures that operate on sets are an under explored area in DL. 

    \item {\em Long Range Dependencies.}
		The DC architecture must be able to determine long range dependencies across attributes and sometimes across relations.
		Many DC tasks rely on the knowledge of such dependencies. 
		They should also be able to leverage additional domain knowledge and integrity constraints.
\end{itemize}

\subsection{Design Space for DC-DL Architectures}
\label{subsec:dlDCArchLangModeling}

Broadly speaking, there are two areas in which DL architectures could be used in DC.
First, novel DL architectures are needed for learning effective representations for downstream DC tasks.
Second, we need DL architectures for common DC tasks such as matching, data repair, imputation, and so on. 
Both are challenging on their own right and require significant innovations.

\subsubsection{Architectures for DC Representation Learning}
Ideally, such as in computer vision, the deep neural networks learn general features in early layers and transition to task specific features in the latter layers.
In the computer vision context, the initial layers learn generic features 
such as edges (which can be used in many tasks) 
while latter layers learn high level concepts such as a cat or dog 
(which could be used for image recognition task).
Models trained on large datasets such as ImageNet could be used as feature extractors followed by specific task specific models.
In other words, the DL models act as a set of Lego bricks that could be put together to perform 
the required functionality. 
It is thus important that the DL architecture for DC follows this property
by learning features that are generic and could readily generalize to many DC tasks.

Achieving the above desirable property for DC is extremely challenging.
Even in an intensively studied area such as NLP, some preliminary breakthroughs happened 
as recently as 2018~\cite{peters2018deep,devlin2018bert,howard2018universal,radford2018improving}.
An ideal representation must be a good approximation for language modeling if it could be used for many NLP tasks. 
Over the last few years, this was primarily done using word embeddings.
However, they are only a crude approximation of language modeling by focusing on co-occurrence.
Due to their limitations, they are used as \emph{inputs} to the first layer of NLP DL architectures that were then trained for specific tasks. 
Recent breakthroughs such as ELMo~\cite{peters2018deep}, ULMFiT~\cite{howard2018universal} and Transformer~\cite{radford2018improving}
have shown how to obtain more expressive language modeling.
The output of these methods are pretrained language models that could be used for various tasks such as 
classification, question answering, semantic matching, and coreference resolution.

While there has been extensive work in linguistics about the hierarchical representations of language, 
a corresponding investigation in data curation and databases is lacking.
Nevertheless, we believe that a generic DC representation learning architecture must be inspired based on the pretrained language models.
Once such an architecture is identified, the learned representations could be used for multiple DC tasks. 

\subsubsection{DL Architectures for DC Tasks}
While the generic approach is often powerful, it is important that we must also work on DL architectures for specific DC tasks.
This approach is often much simpler and provides incremental gains while also increasing our understanding of DL-DC nexus.
In fact, this has been the approach taken by the NLP community - 
they worked on DL architectures specific tasks (such as text classification or question answering) 
even while they searched for more expressive generative language models.

Two recent works, namely DeepMatcher~\cite{Mudgal:2018:DLE:3183713.3196926} and DeepER~\cite{deeper}, provide a template of how this could be achieved.
We discuss them in more details in Section~\ref{subsec:di}.
Both approaches split the task of entity matching into multiple simpler problems and build DL models for them that are then put together. 
Specifically, DeepMatcher split the problem into four components, namely attribute embedding, attribute summarization, attribute similarity and matching.
Once such components are identified, it is possible to use pre-existing DL models for these.
For example, DeepMatcher uses fastText for attribute embedding; a two-layer fully connected ReLU HighwayNet for matching, and so on.

A promising approach is to take specific DC tasks and break them into simpler components and 
evaluate if there are existing DL architectures that could be reused.
Often, it is possible to ``compose'' individual DL models to create a more complex model for solving a specific DC task.
In other words, each of the models are connected such as Lego bricks to form a functional unit. 
Many of the DC tasks could be considered as a combination of tasks such as 
matching, error detection, data repair, imputation, ranking, discovery, and syntactic/semantic transformations.
An intriguing research question is to identify DL models for each of these and investigate how to instantiate and compose them together.

\subsection{Pre-Trained DL Models for DC}
\label{subsec:preTrainedModels}

In domains such as image recognition, there is a common tradition of 
training a DL model on a large dataset and then reusing it (after some tuning) for tasks on other smaller datasets. 
Often, one trains the model on a large, diverse and generic dataset such as ImageNet~\cite{imagenet} (which contains almost 14M images over 20k categories).
The spatial hierarchy learned by this network is often a proxy for modeling the visual world~\cite{chollet2018deep}. 
In other words, many of the features learned by a model trained on ImageNet, such as VGG16, prove useful for many other computer vision problems  -- 
even when it is applied on a different dataset and even different categories~\cite{chollet2018deep}. 
Similarly, in NLP, word embeddings are an effective approximation for language modeling.
The embeddings are often learned from large diverse corpora such as Wikipedia or PubMed and could be used for downstream NLP tasks. 
These pre-trained models can be used in two ways: 
\begin{enumerate}
	\itemsep0em
	\item \emph{feature extraction} where these are used to extract generic features that are fed to a separate classifier for the task at hand; 
	\item \emph{fine-tuning} where one adjusts the abstract representations from the last few hidden layers of a pre-trained model and make it more relevant to a targeted task.
\end{enumerate}

Hence, a promising research avenue is the design of pre-trained DL models for DC that could be used by others with comparatively less resources.

\subsection{Early Successes of DL for DC}
\label{subsec:lowHanging}
In this subsection, we describe the early successes achieved by QCRI and UW-Madison on using DL for major DC tasks such as data discovery and entity matching. 

\subsubsection{Data Discovery}
\label{subsec:dd}

\stitle{Data discovery} is one of the fundamental problems in DC.
Large enterprises typically possess hundreds or thousands of databases and relations.
Data required for analytic tasks is often scattered across multiple databases depending on who collected the data.
This makes the process of finding relevant data for a particular analytic task very challenging.
Usually, there is no domain expert who has complete knowledge about the entire data repository. 
This results in a non-optimal scenario where the data analyst patches together data from her prior knowledge or limited searches 
-- thereby leaving out potentially useful and relevant data.

\stitle{Our First Attempt.}
As an example of applying word embeddings for data discovery,  we show in~\cite{sspaper} how to discover semantic links between  different data units, which are materialized in the \emph{enterprise knowledge graph}  (EKG)\footnote{An EKG is a graph structure whose nodes
are data elements such as tables, attributes, and reference data such as ontologies and
mapping tables, and whose edges represent different relationships between nodes.}.
These links assist in data discovery by linking tables to each other, to
facilitate navigating the schemas, and by relating data to external data
sources such as ontologies and dictionaries, to help explain the schema meaning.
A key component of our work is a semantic matcher based on word embeddings.
It introduces the new concept of coherent groups to tackle the issues of multi-word phrases and out-of-vocabulary terms -- the key idea is that a group of words is similar to another group of words if the average similarity in the embeddings between all pairs of words is high.
In particular,  our approach was able to surface links that were previously unknown to the analysts, \eg 
\emph{isoform}, a type of protein, with \emph{Protein} and \emph{Pcr} -- polymerase chain reaction, a technique to
amplify a segment of DNA -- with assay. 
It also helped discard spurious results obtained from other syntactical and structural matchers, 
\eg the link between \emph{biopsy site} and \emph{site\_components} because the words 
biopsy and components 
do not often appear together -- they are not semantically related. 
More examples on how useful our method is for data discovery including its deployment with a pharmaceutical company can be found in the paper.

\stitle{Ongoing Effort.}
We are trying to use some of the recent advances in Neural Information retrieval~\cite{DBLP:journals/corr/ZhangRBDCKMABKM16} to tackle discovery in the context of DC.
At its core, information retrieval involves two key steps: 
(a) generating good representations for query and documents and 
(b) finding relevance between query and documents.
DL has been applied for both steps.
We can represent appropriate data units (tuples, columns or relations) using either general word embeddings~\cite{mikolov2013distributed} or 
an enterprise specific one that is learned by some of the unsupervised representation learning approaches described previously.

Once we have both of the above steps, we can envision a Google-style search engine 
where the analyst can enter certain textual description of the data that she is looking for.
We convert the query to a distributed representation, 
and use a DL/non-DL model to find the relevance between query and other result units.
Once a set of candidate results (such as relations) are obtained, one can use the  EKG to also simultaneously return other datasets that are thematically related.
The analyst can then use these to make an informed decision.

\subsubsection{Entity Matching}
\label{subsec:di}

\begin{figure}[t!]
	\centering
	\includegraphics[width=\columnwidth]{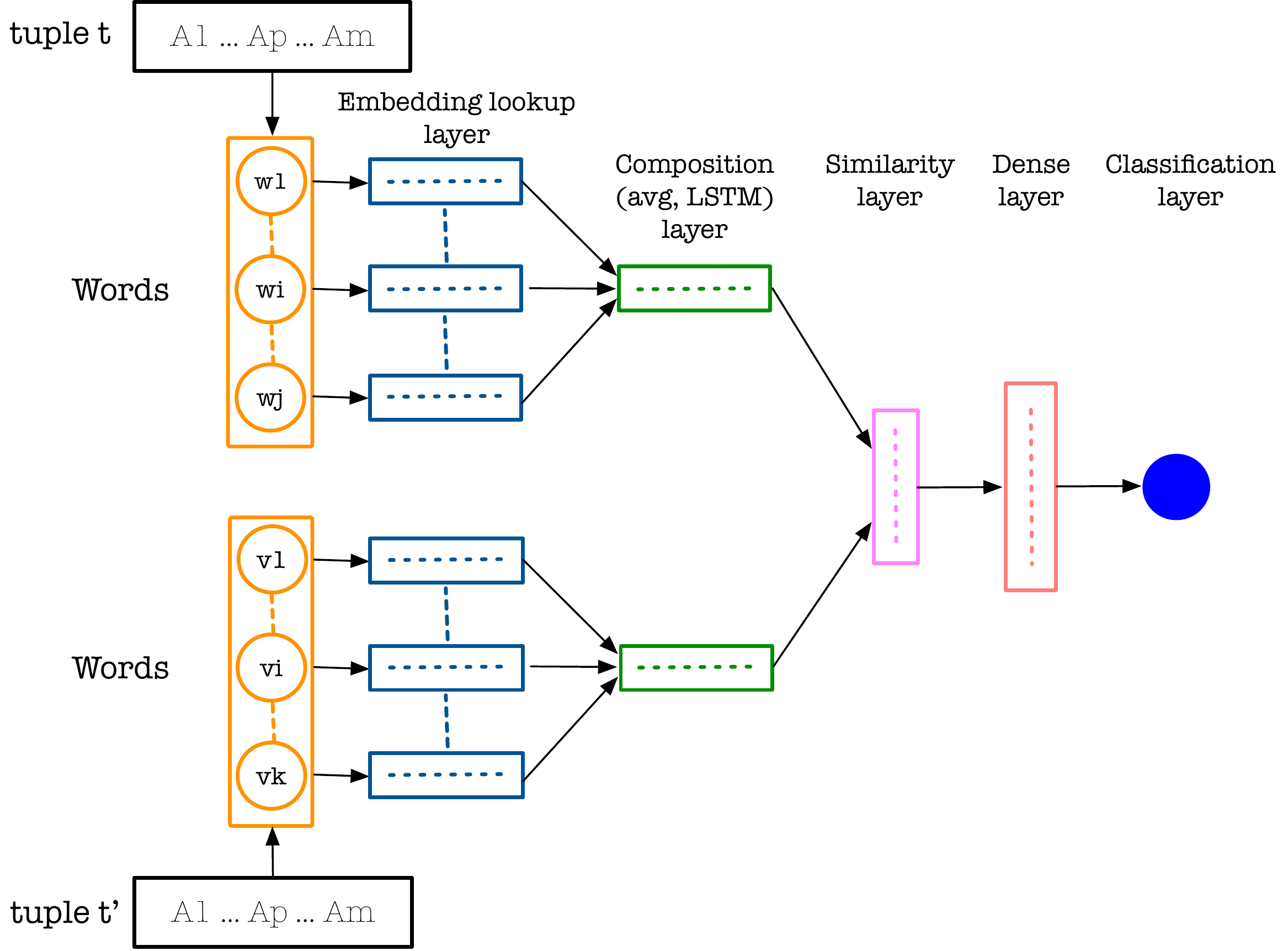}
	\caption{DeepER Framework}
	\label{fig:ERmodel}
\end{figure}

Entity matching is a key problem in data integration, which determines if two tuples refer to the same underlying real-world entity\cite{DBLP:journals/tkde/ElmagarmidIV07}.

\stitle{DeepER.}
Our recent work~\cite{deeper}, DeepER, applies DL techniques for ER.
We show the overall architecture of DeepER in Figure~\ref{fig:ERmodel}.
DeepER pushes the boundaries of existing ER solutions in terms of accuracy, efficiency, and ease-of-use.
For accuracy, we use sophisticated composition methods, namely uni- and bi-directional recurrent neural networks (RNNs) with long short term memory (LSTM) hidden units to convert each tuple  to a distributed representation, 
which can in turn be used to effectively capture similarities between tuples.
For efficiency, we propose a locality sensitive hashing (LSH) based approach that uses distributed representations of tuples for blocking; it takes all attributes of a tuple into consideration and produces much smaller blocks, compared with traditional methods that consider only few attributes.
For ease-of-use, our approach requires much less human labeled data, and does not need feature engineering, compared with traditional machine learning based approaches which require handcrafted features, and similarity functions along with their associated thresholds.
When run on multiple benchmark datasets, DeepER achieves competitive results with minimal interaction with experts.

\begin{figure}[t!]
	\centering
	\includegraphics[width=\columnwidth]{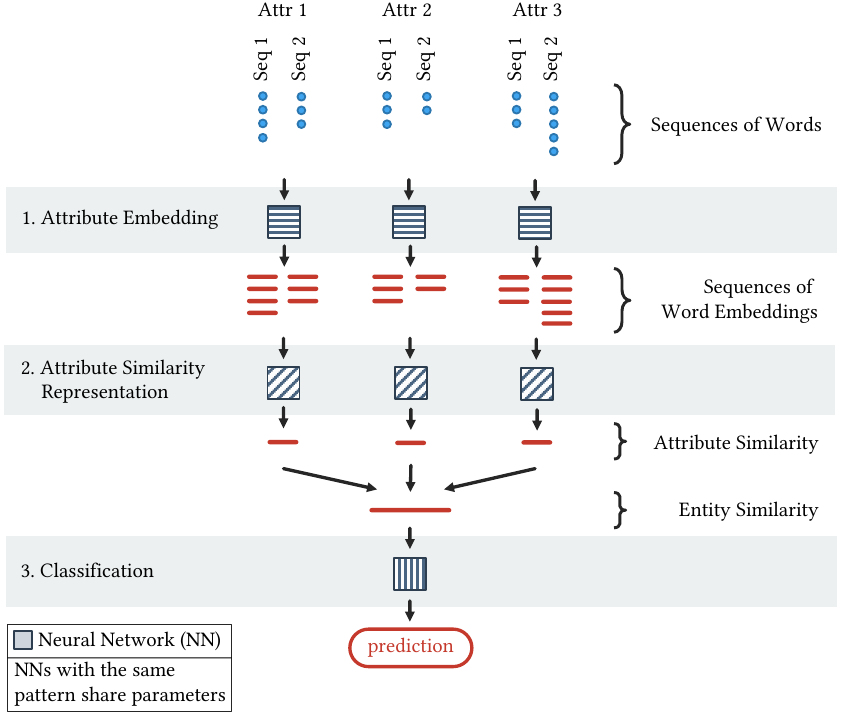}
	\caption{DeepMatcher Framework}
	\label{fig:deepmatcher}
\end{figure}

\stitle{DeepMatcher}~\cite{Mudgal:2018:DLE:3183713.3196926} proposes a template based architecture for entity matching. 
Figure~\ref{fig:deepmatcher} shows an illustration. 
It identifies four major components: attribute embedding, attribute summarization, attribute similarity and matching.
It proposes various choices for each of these components leading to a rich design space. 
The four most promising models differ primarily in how the attribute summarization is performed
and are dubbed as SIF, RNN, Attention, and Hybrid.
SIF model is the simplest and computes a weighted average aggregation over word embeddings to get the attribute embedding.
RNN uses a bi-directional GRU for summarizing an attribute for efficiency.
Attention uses a decomposable attention model that can leverage information from aligned attributes of both tuples to summarize the attribute.
Finally, the hybrid approach combines RNN and attention.
DeepMatcher was evaluated on a diverse set of datasets such as structured, unstructured and noisy and provides useful rule of thumb on when to use DL for ER.

\section{Taming TAC's Hunger for Data}
\label{sec:dldcHunger}

Deep learning often requires a large amount of training data.
Unfortunately, 
in domains such as DC, it is often prohibitively expensive to get high quality training data.
Hence, in order to make \sys a success, \ie using DL to solve DC problems, we must overcome the problem of obtaining training data.
Fortunately, there are a number of exciting principles routinely used in DL that can be adapted to solve this issue. 
Below, we highlight some of these promising ideas and discuss some potential research questions.

\subsection{Unsupervised Representation Learning}

While the amount of \emph{supervised} training data is limited,
most enterprises have substantial amount of \emph{unsupervised} data in various relations and data lakes.
One promising way is to use these unlabeled information to learn some generic patterns from the data.
Once these are identified, one can readily train a DL model with relatively less training data using these representations.
This technique has been successful in NLP where word embeddings are learned on large unlabeled corpus data such as Wikipedia; 
these learned representations are found to provide good performance for many downstream NLP tasks.

\subsubsection*{Research Opportunities}

\begin{itemize}
	\itemsep-1ex 
	\item In the context of DC, we must seek to perform a {\em holistic representation learning} over the enterprise data and 
use the learned representations as features for downstream DC tasks such as entity matching, schema matching, etc. 
\end{itemize}

\newpage
\subsection{Data Augmentation}

Data augmentation is a popular technique in DL to increase the size of labeled training data without increasing the load on domain experts.
The key idea is to apply a series of \emph{label preserving transformations} over the existing training data.
Consider an image recognition task for distinguishing cats from dogs with limited training data.
A typical trick is to apply a series of transformations such as translation (moving the location of the dog/cat within the image),
rotation (changing the orientation), shearing, scaling, changing brightness/color, and so on.
Each of these operations does not change the label of the image -- yet provides many more synthetic training data.

\subsubsection*{Research Opportunities}

\begin{itemize}
	\itemsep0em
    \item {\em Label Preserving Transformations for DC.} 
		What does label preserving transformations mean for DC? 
		Is it possible to design an algebra of such transformations? 
    \item {\em Domain Knowledge Aware Augmentation.}
    	To avoid creating erroneous data, we could integrate domain knowledge and integrity constraints in the data augmentation process.
   		This would ensure that we do not create tuples that say New York City is the capital of USA.  
    \item {\em Domain Specific Transformations.} 
    	There are some recent approaches such as Tanda~\cite{tanda} that seek to learn domain specific transformations.
		For example, if one knows that \emph{Country} $\rightarrow$ \emph{Capital}, 
		we can just swap the (Country, Capital) values of two tuples to generate two new tuples. 
		In this case, even if the new tuple is not necessarily real, its label will be the same as the original tuple.
		Is it possible to develop an algebra for such transformations? 
\end{itemize}

\subsection{Synthetic Data Generation for DC} 
A seminal event in computer vision was the construction of ImageNet dataset~\cite{imagenet} with many million images over thousands of categories.
This dataset was orders of magnitude larger than prior ones and served as a benchmark for image recognition for many years.
This also allowed the development of powerful DL algorithms.
We believe that the DC community is in need of such an ImageNet moment. 

\subsubsection*{Research Opportunities}

\begin{itemize}
	\itemsep0em
    \item {\em Benchmark for DC.}
    	It is paramount to create a similar benchmark to drive research in DC (both DL and non-DL).
        While there has been some work for data cleaning such as BART~\cite{bart}, it is often limited to specific scenarios.
        For example, BART can be used to benchmark data repair algorithms where the integrity constraints are specified as denial constraints.
    \item {\em Synthetic Datasets.}
		If it is not possible to create an open-source dataset that has realistic data quality issues, 
        a useful fall back is to create synthetic datasets that exhibit representative and realistic data quality issues.
        The family of TPC benchmarks involves a series of synthetic datasets that is somehow realistic and widely used for benchmarking database systems.
        How can one use a wide variety of DL techniques to generate synthetic data?
        The most promising approaches are variational auto encoders (VAE) and Generative adversarial networks (GANs).
        However, they both have their own pros and cons.
        While the latent space of VAE is more structured, it also makes additional distributional assumptions that might not strictly apply to DC.
        GANs on the other hand are more generic but often have issues with convergence.
\end{itemize}

\subsection{Weak Supervision}
A key bottleneck in creating training data is that there is often an implicit assumption that it must be accurate.
However, it is often infeasible to produce sufficient hand-labeled and accurate training data for most DC tasks.
This is especially challenging for DL models that require a huge amount of training data.
However, if one can relax the need for the veracity of training data, its generation will become much easier for the expert.
The domain expert can specify a high level mechanism to generate training data without endeavoring to make it perfect.
For example, she can say that if two tuples have the same country but different capitals, they are erroneous.
This often requires substantially less effort than manually encoding all the exceptions.

\subsubsection*{Research Opportunities}
\begin{itemize}
	\itemsep0em
    \item {\em Weakly Supervised DL Models.}
    	There has been a series of research (such as Snorkel~\cite{ratner2017snorkel}) that seek to 
		\emph{weakly supervise} ML models and provide a convenient programming mechanism to specify ``mostly correct'' training data. 
		What are the DC specific mechanisms and abstractions for weak supervision? 
		Can we automatically create such a weak supervision through data profiling? 
\end{itemize}

\subsection{Domain Adaptation} 

Another trick to handle limited training data is to ``transfer'' representations learned in one task to a different yet related task.
For example, DL models for computer vision tasks are often trained on ImageNet~\cite{imagenet},  
a dataset that is commonly used for image recognition purposes.
However, these models could be used for tasks that are not necessarily image recognition 
and are even used for recognizing images that do not belong to the categories found in ImageNet.
This approach of training on a large diverse dataset followed by tuning for a local dataset and tasks has been very successful.

\subsubsection*{Research Opportunities}
\begin{itemize}
	\itemsep0em
    \item {\em Transfer learning.} 
		What is the equivalent of this approach for DC?
		Is it possible to train on a single large dataset such that it could be used for many downstream DC tasks?
		Alternatively, is it possible to train on a single dataset and for a single task (such as entity matching)
		such that the learned representations could be used for entity matching in similar domains?
    	How can we train a DL model for one task and tune the model for new tasks by using the limited labeled data instead of starting from scratch?
    \item {\em Pre-trained Models.} 
		Is it possible to provide pre-trained models that have been trained on large, generic datasets?
		These models could then be tuned by individual practitioners in different enterprises.
\end{itemize}

\subsection{Crowdsourcing}

Recently, there has been increasing interest in using crowdsourcing for a variety of tasks 
including generating appropriate training data for ML models.
The output of crowd workers is often noisy and hence requires sophisticated algorithms~\cite{li2017crowdsourced} for 
inferring true labels from noisy labels, learning the skill of workers, assigning workers to appropriate tasks, and so on.

\subsubsection*{Research Opportunities}

\begin{itemize}
    \item {\em Crowdsourced DL.} 
		Combining crowdsourcing along with some of the aforementioned ideas (such as data augmentation) is an intriguing research problem to investigate.
		How does the noisiness of the crowdsourced labels interact with the inherent uncertainty of weak supervision?
		Can crowd workers come up with novel data augmentation strategies? What is an appropriate mechanism to elicit them?
\end{itemize}

\section{Deep Learning: Myths and Concerns}
\label{sec:discussion}

In the past five years, DL has achieved substantial successes in many areas such as computer vision, NLP, speech recognition, and many more.
As pointed out in the previous sections, DC has a number of characteristics that are quite different from prior domains where DL succeeded.
We anticipate that applying DL to challenging real-world DC tasks can be messy.
We now describe some of the concerns that could be raised by a pragmatic DC practitioner or a skeptical researcher.

\subsection{Deep Learning is Computing Heavy}

A common concern is that training DL models requires exorbitant computing resources.
There are legions of tales on how training some DL models took days even on a large GPU cluster.
In practice, training time often depends on the model complexity, such as the number of parameters to be learnt, and the size of training data.

There are many tricks that can reduce the amount of training time.
For example, a task-aware DL architecture often requires substantially less parameters to be learned.
Alternatively, one can ``transfer'' knowledge from a pre-trained model from a related task or domain 
and the training time will now be proportional to the amount of fine-tuning required to customize the model to the new task. 
For example, DeepER~\cite{deeper} leveraged word embeddings from GloVe (whose training can be time consuming)
and built a light-weight DL model that can be trained in a matter of minutes even on a CPU.
Finally, while training could be expensive, this can often be done as a pre-processing task.
Prediction using DL is often very fast and  comparable to that of other ML models.

\subsection{Data Curation Tasks are Too Messy or Too Unique}

\stitle{DC tasks often require substantial domain knowledge and a large dose of ``common sense''.}
Current DL models are very narrow in the sense that they primarily learn from the correlations present in the training data.
However, it is quite likely that this might not be sufficient.
Unsupervised representation learning over the entire enterprise data can only partially address this issue.
Current DL models are often not amenable to encoding domain knowledge in general as well as those that are specific to DC such as data integrity constraints.
As mentioned before, substantial amount of new research on DC-aware DL architectures is needed.
However, it is likely that DL, even in its current form, can reduce the work of domain experts.

\stitle{DC tasks often exhibit a \emph{skewed label distribution}.}
For the task of entity resolution (ER), the number of non-duplicate tuple pairs are orders of magnitude larger than the number of duplicate tuple pairs.
If one is not careful, DL models can provide inaccurate results.
Similarly, other DC tasks often exhibit \emph{unbalanced cost model} where the cost of misclassification is not symmetric.
Prior DL work utilizes a number of techniques to address these issues such as
(a)~cost sensitive models where the asymmetric misclassification costs are encoded into the objective function, and 
(b)~sophisticated sampling approach where we under or over sample certain classes of data.
For example, DeepER~\cite{deeper} samples non-duplicate tuple pairs that are abundant at a higher level than duplicate tuple pairs.

\subsection{Deep Learning Predictions are Inscrutable}
Yet another concern from domain experts is that the predictions of DL models are often uninterpretable.
Deep learning models are often very complex and the black-box predictions might not be explainable by even DL experts.
However, explaining why a DL model made a particular data repair is very important for a domain expert.
Recently, there has been intense interest in developing algorithms for explaining predictions of DL models 
or designing interpretable DL models in general.
Please see~\cite{guidotti2018survey} for an extensive survey.
Designing algorithms that can explain the prediction of DL models for DC tasks is an intriguing research problem.

\subsection{Deep Learning can be Easily Fooled}

There exist a series of recent works which show that DL models (especially for image recognition) 
can be easily fooled by perturbing the images in an adversarial manner.
The sub-field of adversarial DL~\cite{szegedy2013intriguing,papernot2016limitations} studies 
the problem of constructing synthetic examples by slightly modifying real examples from training data 
such that the trained DL model (or any ML model) makes an incorrect prediction with high confidence.
While this is indeed a long term concern, most DC tasks are often collaborative and limited to an enterprise.
Furthermore, there are a series of research efforts that propose DL models that are more resistant to adversarial training such as~\cite{madry2017towards}. 

\subsection{Building Deep Learning Models for Data Curation is ``Just Engineering''}
Many DC researchers might look at the process of building DL models for DC and simply dismiss it 
as a pure engineering effort.
And they are indeed correct!
Despite its stunning success, DL is still at its infancy and the theory of DL is still being developed.
To a DC researcher used to purveying a well organized garden of conferences such as VLDB/SIGMOD/ICDE,
the DL research landscape might look like the wild west. 

In the early stages, researchers might \emph{just apply} an existing DL model or algorithm for a DC task.
Or they might slightly tweak a previous model to achieve better results.
We argue that database conferences must provide a safe zone in which these DL explorations are conducted in a principled manner.
One could take inspiration from the computer vision community.
They created a large dataset (ImageNet~\cite{imagenet}) that provided a benchmark by which different DL architectures and algorithms can be compared.
They also created one or more workshops focused on applying DL for specific tasks in computer vision 
(such as image recognition and scene understanding).
The database community has its own TPC series of synthetic datasets that have been influential in benchmarking database systems.
Efforts similar to TPC are essential for the success of DL-driven DC.

\subsection{Deep Learning is Data Hungry}
Indeed, this is one of the major issues in adopting DL for DC.
Most classical DL architectures often have many hidden layers with millions of parameters to learn, which requires a large amount of training data.
Unfortunately, the amount of training data for DC is often small.
The good news is, there exist a wide variety of techniques and tricks in the DL's arsenal that can help address this issue. Below, we summarize the most promising directions (some were discussed in Section~\ref{sec:dldcHunger}):  
\begin{itemize}
	\item leveraging the vast amount of unlabeled data to learn key parameters;
	\item transferring knowledge gained from a related task/domain;
	\item designing DC-aware DL architectures that require less data to train; and
	\item obtaining cheap but possibly inaccurate labels instead of expensive but accurate labels.
\end{itemize}

\section{A Call to Arms}
\label{sec:conclusion}

In this paper, we presented our vision to build \sys, an ecosystem to unleash the power of DL in DC.
This vision is motivated by two key observations.
Data Curation is a long standing problem and needs novel solutions in order to handle the emerging big data ecosystem.
Deep Learning is gaining traction across many disciplines, both inside and outside computer science,
showing successes, which were unthinkable, just a few years ago.
The meeting of these two disciplines will unleash a series of research activities that will certainly lead
to actual and usable solutions for many DC tasks.

Key components of our roadmap in \sys include research opportunities in 
learning distributed representations for database aware objects such as tuples or columns and designing DC-aware DL architectures.
We described a number of promising approaches to tame DL's hunger for data. 
We discuss the early successes in using DL for important DC tasks such as data discovery and entity matching 
that required novel adaptations of DL techniques.
Moreover, we have described many anecdotal tricks for DL learned from other domains that are likely to be useful for DC.
All in all, besides presenting our vision and early achievement of \sys, this is a call to arms for the database community in general, and the DC community in particular, to seize this opportunity to significantly advance the area, while keeping in mind the risks and mitigation that were also highlighted in this paper.

\bibliographystyle{abbrv}
\bibliography{DA}

\balance

\end{document}